**Temperature–doping phase diagram and endurance in Ce-doped $HfO_2$**

Amit Kumar Shah,[1*] Haidong Lu,[1*] Kawshan Hathurusingha,[1*] Alexei Gruverman,[1,2,a)] Xiaoshan Xu[1,2,a)]

[1]Department of Physics and Astronomy, University of Nebraska, Lincoln, Nebraska 68588, USA

[2]Nebraska Center for Materials and Nanoscience, University of Nebraska, Lincoln, Nebraska 68588, USA

[*]These authors contributed equally to this work.

a) Author to whom correspondence should be addressed: agruverman2@unl.edu, xiaoshan.xu@unl.edu

**Abstract**

The structural and ferroelectric properties of epitaxial $Hf_{1-x}Ce_xO_2$ (CHO) thin films in the ultrathin regime are investigated as a function of Ce concentration ($5\% \leq x \leq 20\%$) and temperature. A temperature–doping phase diagram is established for 10 nm films, showing a systematic evolution from the ferroelectric orthorhombic phase to tetragonal and cubic phases with increasing Ce content. The orthorhombic–tetragonal transition temperature decreases from ~800 °C at $x = 5\%$ to ~300 °C at $x = 15\%$, indicating strong stabilization of higher-symmetry phases with doping. Consistently, the remanent polarization decreases from ~15 to 3.8 μC/cm$^2$ as $x$ increases from 5% to 20%. In contrast, the endurance improves significantly, with higher Ce concentrations exhibiting markedly enhanced cycling stability up to $10^8$ cycles. The opposing trends of polarization and endurance are correlated with reduced orthorhombic distortion, suggesting that fatigue mitigation in Ce-doped $HfO_2$ is linked to structural evolution. These results provide a framework for optimizing composition and reliability in ultrathin ferroelectric $HfO_2$ devices.

## Introduction

Hafnium oxide–based ferroelectrics have attracted significant attention for microelectronic applications due to their compatibility with silicon technology and the persistence of ferroelectricity at nanometer length scales.

A central challenge is the stabilization of the ferroelectric orthorhombic phase (o-FE, space group *$Pca2_1$*), as the thermodynamically stable polymorph of $HfO_2$ under ambient conditions is the nonpolar monoclinic phase (m, space group *$P2_1/c$*). The o-FE phase is closely related to a high-pressure orthorhombic phase (*Pbca*) and shares the tetragonal phase (t, space group *$P4_2/nmc$*) as a common parent phase with the monoclinic structure in the bulk phase diagram.[1] Various strategies have been explored to stabilize the o-FE phase during fabrication, including strain engineering,[2] control of oxygen vacancy concentration,[3] surface and interface energy effects,[4,5] and the kinetic suppression of the tetragonal-to-monoclinic transformation, which typically exhibits a larger energy barrier than the transformation from tetragonal to orthorhombic phases.

In addition to phase stabilization during processing, maintaining the o-FE phase under electrical cycling is critical for device reliability. Polarization switching can induce a gradual conversion from the orthorhombic to the monoclinic phase, which is considered a major contributor to fatigue in hafnia-based ferroelectrics.[6,7] In particular, the accumulation and redistribution of oxygen vacancies significantly increases the probability of such phase transformation during repeated switching.[6,7] Ce doping has been proposed as an effective approach to mitigate vacancy-related degradation, as the mixed $Ce^{3+}/Ce^{4+}$ valence can impede oxygen vacancy migration and accumulation.[8–11]

Despite these advances, the role of Ce doping in governing phase formation in $HfO_2$ remains insufficiently understood.[12,13] Although $HfO_2$ and $CeO_2$ both belong to fluorite-structure-related oxide family, the equilibrium phase diagram of the $HfO_2$–$CeO_2$ system exhibits limited solid solubility,[14] with only a few percent dopant incorporation achievable without substantially modifying the host lattice structure. This limited miscibility complicates efforts to systematically study the structural impact of Ce incorporation in ferroelectric $HfO_2$ thin films. Previous studies on Ce-doped $HfO_2$ have primarily focused on films processed by rapid thermal annealing (RTA),[10–12,15–20] where phase formation is strongly influenced by nonequilibrium kinetics. As a result, the reported dependence of structural stability on doping concentration varies significantly across studies.[11,12,15,17] Furthermore, most investigations have been performed on relatively thick films (20–200 nm),[10–12,15–20] whereas ferroelectric $HfO_2$ exhibits fundamentally different behavior in the ultrathin regime, where enhanced ferroelectricity and altered phase stability are observed. While some reports have examined switching endurance[10,11] and others have probed structural evolution,[11,12,15,17] a systematic correlation between doping-dependent phase stability and endurance remains limited.

In this work, we investigate Ce-doped epitaxial $Hf_{1-x}Ce_xO_2$ (CHO) thin films with Ce concentrations ranging from 5% to 20%, in the ultrathin regime (4–10 nm), where ferroelectric behavior is expected to be enhanced. Unlike prior studies based on rapid thermal annealing, the films are grown at elevated temperature under conditions closer to thermodynamic equilibrium, enabling a more intrinsic evaluation of doping-dependent phase stability. By focusing on a fixed thickness of 10 nm, we construct a temperature–doping structural phase diagram that captures the evolution among orthorhombic, tetragonal, and cubic phases. We find that increasing Ce concentration systematically drives the crystal structure toward higher symmetry, accompanied by a reduction in the orthorhombic–tetragonal transition temperature and a decrease in polarization. Furthermore, by correlating structural evolution with polarization switching and endurance measurements, we show that reduced orthorhombic distortion is associated with enhanced endurance, suggesting that the fatigue mitigation induced by Ce doping has an intrinsic structural origin.

## Experimental

Epitaxial Ce-doped $HfO_2$ (CHO, $Hf_{1-x}Ce_xO_2$) thin films with thicknesses in the range of 3–15 nm were grown by pulsed laser deposition (PLD) on $SrTiO_3$ (STO) (110) substrates buffered with a 15–20 nm $La_{0.7}Sr_{0.3}MnO_3$ (LSMO) bottom electrode at about 750 °C in 70 mTorr $O_2$. The ablation was done using a KrF Excimer laser with wavelength 248 nm at a 2 Hz repetition rate. The Ce concentration $x$ was varied between 5% and 20%. Unless otherwise noted, the discussion below focuses on films with a thickness of ~10 nm. For electrical measurement, Pt top electrodes with a thickness of ~15 nm and a diameter ~10 μm were fabricated using vacuum deposition and photo lithography.

Reflection high energy electron diffraction (RHEED) built-in the growth chamber was used to study the temperature dependence of the structure of the CHO films during warming. The temperature of the films was measured using an external pyrometer. X-ray diffraction θ–2θ scans and X-ray reflectivity (for film thickness), were measured with a Rigaku SmartLab diffractometer using Cu Kα radiation (1.54 Å). The specular and non-specular pseudo cubic (111) planes as well as the pseudo cubic (004) and (110) planes of the CHO films were measured with a Bruker AXS D8 Discover diffractometer using an area detector (1.54 Å).

Polarization switching was measured using a PUND (positive up negative down) setup, where two consecutive positive (or negative) triangular wave pulses at a frequency of 100 kHz generated by a Keysight 33621A arbitrary waveform generator were applied to the Pt top electrode, and the resulting switching current from the capacitor was collected via a Tektronix TDS 3014B oscilloscope with a current to voltage THS4021 EVM amplifier. The parasitic non-switching current was removed by subtraction of the second pulse signal from the first.

## Temperature-doping phase diagram for 10 nm films

**Figure 1a** shows specular x-ray diffraction (XRD) patterns for CHO films with varying Ce concentration. A dominant Bragg peak is observed near 30°, which can be indexed to either the orthorhombic (111) or tetragonal (110) reflection. Pronounced Laue oscillations are present for all compositions, indicating high crystalline quality and atomically smooth interfaces. Notably, no reflections associated with the monoclinic phase (typically near 28° and 31°) are detected, consistent with previous reports of doped $HfO_2$ films grown on LSMO (110).[21,22] The suppression of the monoclinic phase can be attributed to epitaxial growth along the pseudocubic (111) direction under compressive strain, which destabilizes the monoclinic structure.[2,23]

To distinguish between the orthorhombic (o) and tetragonal (t) phases at room temperature, reflection high-energy electron diffraction (RHEED) measurements were performed. Representative RHEED patterns for ~10 nm films are shown in **Fig. 1b**, with the electron beam aligned along the LSMO $[1\bar{1}0]$ direction. Diffraction peaks corresponding to $(h, -h, 0)$ reflections in pseudocubic notation are observed. While reflections with even $h$ are allowed for orthorhombic, tetragonal, and cubic phases, those with odd $h$ are only allowed for the orthorhombic phase. Therefore, the intensity ratio between the $(1\bar{1}0)$ and $(2\bar{2}0)$ reflections serves as a structural order parameter for the orthorhombic phase.[24,25]

As the Ce concentration increases, the $(1\bar{1}0)$ peak intensity decreases systematically, while the $(2\bar{2}0)$ peak remains nearly unchanged, indicating a reduction in orthorhombic phase fraction and a transition toward higher-symmetry phases. At $x \approx 20\%$, the $(1\bar{1}0)$ peak becomes barely discernible, suggesting that the system approaches the orthorhombic–tetragonal phase boundary at room temperature. This trend is summarized in **Fig. 1c**, where the intensity ratio decreases monotonically with increasing $x$, indicating that the orthorhombic phase becomes unstable at compositions slightly above 20%.

In addition to the orthorhombic-to-tetragonal transition, the emergence of a cubic phase at higher Ce concentrations is expected. To probe the tetragonal-to-cubic transition, we analyzed the unit cell symmetry using high-angle XRD measurements near the (400) reflections (**Fig. 2a**). For $x = 5\%$, two distinct peaks are resolved, corresponding to anisotropic lattice parameters consistent with orthorhombic (or weakly distorted tetragonal) symmetry. As the Ce concentration increases, the lattice constants evolve anisotropically: the smaller lattice parameter increases more rapidly than the larger ones, resulting in a progressive reduction of orthorhombic distortion. This evolution is summarized in **Fig. 2b** and indicates a continuous approach toward cubic symmetry. Extrapolation of the lattice parameters suggests convergence at $x \sim 27\%$, consistent with stabilization of the cubic fluorite structure at high Ce content.

Having established the doping-induced phase evolution at room temperature, we next examine the effect of temperature on phase stability. The orthorhombic-to-tetragonal transition temperature ($T_{ot}$) was determined from the temperature dependence of the RHEED intensity ratio. **Figure 3a** shows a representative measurement for $x = 15\%$, where the order parameter decreases with increasing temperature and vanishes at $T_{ot}$.

As summarized in **Fig. 3b**, $T_{ot}$ decreases dramatically with increasing Ce concentration, from ~800 °C at $x$ = 5% to ~300 °C at $x$ = 15%. This strong suppression of $T_{ot}$ indicates that Ce doping stabilizes higher-symmetry phases, consistent with the room-temperature structural evolution discussed above.

Combining the results from **Figs. 1c**, **2b**, and **3b**, we construct the temperature–composition ($x$–$T$) phase diagram for ~10 nm CHO films (**Fig. 3c**), where the tetragonal-to-cubic transition temperature at $x$=0 is the bulk value.[26] The diagram reveals that both increasing temperature and increasing Ce concentration systematically drive the system from the ferroelectric orthorhombic phase toward tetragonal and ultimately cubic phases. Notably, no phase separation is observed across the investigated composition range, highlighting the stabilization of a metastable solid solution under epitaxial constraints, in contrast to bulk equilibrium behavior.

This phase diagram defines the compositional and thermal window for stabilization of the ferroelectric orthorhombic phase, which will be directly correlated with polarization switching and endurance behavior in the following sections.

## Electric polarization and endurance

Polarization switching behavior of 10 nm CHO films with different Ce concentrations is shown in **Fig. 4a**. The extracted remanent polarization ($P_r$) is summarized in **Fig. 4b**. For $x$ = 5%, $P_r$ is approximately 15 μC/cm$^2$, and it decreases systematically with increasing Ce concentration, reaching 3.8 μC/cm$^2$ at $x$ = 20%.

This reduction in $P_r$ with increasing Ce content is consistent with the structural evolution observed in **Fig. 1c**. In particular, the decrease in the RHEED intensity ratio between the $(1\bar{1}0)$ and $(2\bar{2}0)$ reflections indicates a reduction in orthorhombic distortion, which is the structural origin of ferroelectric polarization in $HfO_2$. The continuous suppression of the orthorhombic phase fraction therefore leads directly to the observed decrease in remanent polarization.

The endurance of CHO films was evaluated under repeated polarization switching cycles. **Figure 4c** shows representative polarization–electric field loops for a 10 nm film with $x$ = 15% after different numbers of switching cycles, and the corresponding evolution of $P_r$ is summarized in **Fig. 4d**, together with data for $x$ = 5% and $x$ = 10%. In all cases, $P_r$ decreases with increasing cycling, although an initial increase is observed for $x$ = 15%, which can be attributed to a wake-up effect.

A pronounced dependence of endurance on Ce concentration is observed. For $x$ = 5%, $P_r$ decays rapidly and nearly vanishes after $10^6$ cycles. In contrast, for $x$ = 15%, $P_r$ remains slightly higher than its initial value after wake-up and retains approximately 40% of its initial value even after $10^8$ cycles. The $x$ = 10% sample exhibits intermediate behavior, with a more monotonic and relatively faster decay compared to $x$ = 15%. These results indicate a clear enhancement of endurance with increasing Ce concentration.

This improvement in endurance is consistent with the proposed role of Ce doping in modifying oxygen vacancy behavior. The mixed $Ce^{3+}/Ce^{4+}$ valence can impede the migration and accumulation of oxygen vacancies, thereby suppressing vacancy-driven degradation. In the Pt/CHO/LSMO device structure used here, oxygen vacancies are expected to accumulate preferentially near the Pt/CHO interface due to the chemical inertness of Pt. The resulting local increase in vacancy concentration facilitates the transformation from the orthorhombic ferroelectric phase to the nonpolar monoclinic phase,[27] which reduces polarization and leads to fatigue.

Combined with the structural results, we observe a correlation between reduced orthorhombic distortion and improved cycling stability, which suggests that, in addition to defect-mediated effects, fatigue mitigation in Ce-doped $HfO_2$ may benefit from the reduced structural deformation associated with polarization switching. Besides the commonly invoked role of Ce in reducing oxygen-vacancy mobility,[8–11] the improved endurance may also be related to the reduced structural deformation accompanying polarization switching. Recent work has shown that in $ZrO_2$ films with smaller switching-induced volume changes can display significantly improved endurance.[28] Consistent with this picture, the orthorhombic distortion in CHO decreases with increasing Ce concentration, coinciding with the observed enhancement in cycling stability.

## Conclusion

In summary, we have established a temperature–doping structural phase diagram for ultrathin Ce-doped $HfO_2$ thin films at a thickness of 10 nm. Increasing Ce concentration systematically stabilizes higher-symmetry phases, leading to a reduction in the orthorhombic–tetragonal transition temperature and a corresponding decrease in remanent polarization. Despite this reduction, a pronounced enhancement in endurance is observed at higher doping levels. The correlation between reduced orthorhombic distortion and improved cycling stability indicates that fatigue mitigation in Ce-doped $HfO_2$ is intrinsically linked to structural evolution. These results provide insight into phase stability and offer guidance for balancing ferroelectric performance and reliability in ultrathin $HfO_2$-based devices.

## Acknowledgement

This work is supported by the National Science Foundation (grant DMR-2419172). The research was performed in part at the Nebraska Nanoscale Facility, National Nanotechnology Coordinated Infrastructure and the Nebraska Center for Materials and Nanoscience, which are supported by the NSF under Grant No. ECCS-2025298, as well as the Nebraska Research Initiative through the Nebraska Center for Materials and Nanoscience and the Nanoengineering Research Core Facility at the University of Nebraska–Lincoln.

## References

1. Ohtaka, O. *et al.* Phase Relations and Volume Changes of Hafnia under High Pressure and High Temperature. *Journal of the American Ceramic Society* **84**, 1369–1373 (2001).
2. Liu, S. & Hanrahan, B. M. Effects of growth orientations and epitaxial strains on phase stability of HfO2 thin films. *Physical Review Materials* **3**, 054404 (2019).
3. Ma, L. Y. & Liu, S. Structural Polymorphism Kinetics Promoted by Charged Oxygen Vacancies in $HfO_2$. *Physical Review Letters* **130**, 096801 (2023).
4. Materlik, R., Künneth, C. & Kersch, A. The origin of ferroelectricity in $Hf_{1-x}Zr_xO_2$: A computational investigation and a surface energy model. *Journal of Applied Physics* **117**, 134109 (2015).
5. Batra, R., Tran, H. D. & Ramprasad, R. Stabilization of metastable phases in hafnia owing to surface energy effects. *Applied Physics Letters* **108**, 172902 (2016).
6. Zhang, Z. *et al.* Phase Transformation Driven by Oxygen Vacancy Redistribution as the Mechanism of Ferroelectric $Hf_{0.5}$ $Zr_{0.5}$ $O_2$ Fatigue. *Adv Elect Materials* **10**, 2300877 (2024).
7. Nukala, P. *et al.* Reversible oxygen migration and phase transitions in hafnia-based ferroelectric devices. *Science* **372**, 630–635 (2021).
8. Zhou, C. *et al.* Fatigue-free ferroelectricity in Hf0.5Zr0.5O2 ultrathin films via interfacial design. *Nat Commun* **16**, 7593 (2025).
9. Yu, Z. *et al.* $CeO_2$ Doping of $Hf_{0.5}$ $Zr_{0.5}$ $O_2$ Thin Films for High Endurance Ferroelectric Memories. *Adv Elect Materials* **8**, 2101258 (2022).
10. Shiraishi, T., Konno, T. J. & Funakubo, H. Ferroelectric and piezoelectric properties of 100 nm-thick CeO2-HfO2 epitaxial films. *Applied Physics Letters* **120**, 132901 (2022).

11. Xu, J. *et al.* Excellent Ferroelectricity and Thermal Stability in Ce-Doped HfO2 Thin Films via Oxygen Vacancy Modulation. *ACS Appl. Mater. Interfaces* **18**, 37037–37046 (2026).

12. Shiraishi, T. *et al.* Formation of the orthorhombic phase in CeO2-HfO2 solid solution epitaxial thin films and their ferroelectric properties. *Applied Physics Letters* **114**, 232902 (2019).

13. Tian, Y., Zhou, Y., Zhao, M., Ouyang, Y. & Tao, X. Effect of Ce doping on ferroelectric HfO2 from first-principles: Implications for ferroelectric thin films and phase regulation. *Journal of Solid State Chemistry* **328**, 124316 (2023).

14. Andrievskaya, E. R., Gerasimyuk, G. I., Kornienko, O. A., Samelyuk, A. V. & Lopato, L. M. Phase equilibria in the HfO2-ZrO2-CeO2 system at 1250°C. *Inorg Mater* **42**, 1352–1359 (2006).

15. Zheng, S. *et al.* Improvement of remanent polarization of CeO2–HfO2 solid solution thin films on Si substrates by chemical solution deposition. *Applied Physics Letters* **117**, 212904 (2020).

16. Cüppers, F., Hirai, K. & Funakubo, H. On the switching dynamics of epitaxial ferroelectric CeO2–HfO2 thin film capacitors. *Nano Convergence* **9**, 56 (2022).

17. Hirai, K. *et al.* Composition dependence of ferroelectric properties in (111)-oriented epitaxial $HfO_2$ -$CeO_2$ solid solution films. *Jpn. J. Appl. Phys.* **61**, SN1019 (2022).

18. Shiraishi, T., Choi, S., Kiguchi, T. & Konno, T. J. Structural evolution of epitaxial CeO2-HfO2 thin films using atomic-scale observation: Formation of ferroelectric phase and domain structure. *Acta Materialia* **235**, 118091 (2022).

19. Pang, J. *et al.* The enhanced ferroelectric properties of flexible Hf0.85Ce0.15O2 thin films based on in situ stress regulation. *npj Flex Electron* **9**, 7 (2025).

20. Li, S. *et al.* Enhancing the ferroelectricity of highly (001) oriented lanthanum-doped hafnium oxide films prepared by pulsed laser deposition. *Nano Research* **19**, 94908581 (2026).

21. Yun, Y. *et al.* Intrinsic ferroelectricity in Y-doped $HfO_2$ thin films. *Nature Materials* **21**, 903–909 (2022).

22. Jiao, P. *et al.* Flexoelectricity-stabilized ferroelectric phase with enhanced reliability in ultrathin La:HfO2 films. *Applied Physics Reviews* **10**, 031417 (2023).

23. Zhu, T., Deng, S. & Liu, S. Epitaxial ferroelectric hafnia stabilized by symmetry constraints. *Physical Review B* **108**, L060102 (2023).

24. Mimura, T. *et al.* Effects of heat treatment and in situ high-temperature X-ray diffraction study on the formation of ferroelectric epitaxial Y-doped HfO2 film. *Japanese Journal of Applied Physics* **58**, SBBB09 (2019).

25. Tashiro, Y., Shimizu, T., Mimura, T. & Funakubo, H. Comprehensive Study on the Kinetic Formation of the Orthorhombic Ferroelectric Phase in Epitaxial Y-Doped Ferroelectric HfO2 Thin Films. *ACS Applied Electronic Materials* **3**, 3123–3130 (2021).

26. Wang, J., Li, H. P. & Stevens, R. Hafnia and hafnia-toughened ceramics. *Journal of Materials Science* **27**, 5397–5430 (1992).

27. Paul, T. K., Saha, A. K. & Gupta, S. K. Oxygen vacancy-induced monoclinic dead layers in ferroelectric HfO2 with metal electrodes. *Journal of Applied Physics* **137**, 144102 (2025).

28. Hsieh, H.-Y. *et al.* In-situ TEM analysis of reversible antiferroelectricity in ZrO2 thin films via near-constant-volume tetragonal symmetry transition. *Materials Today* **94**, 103257 (2026).

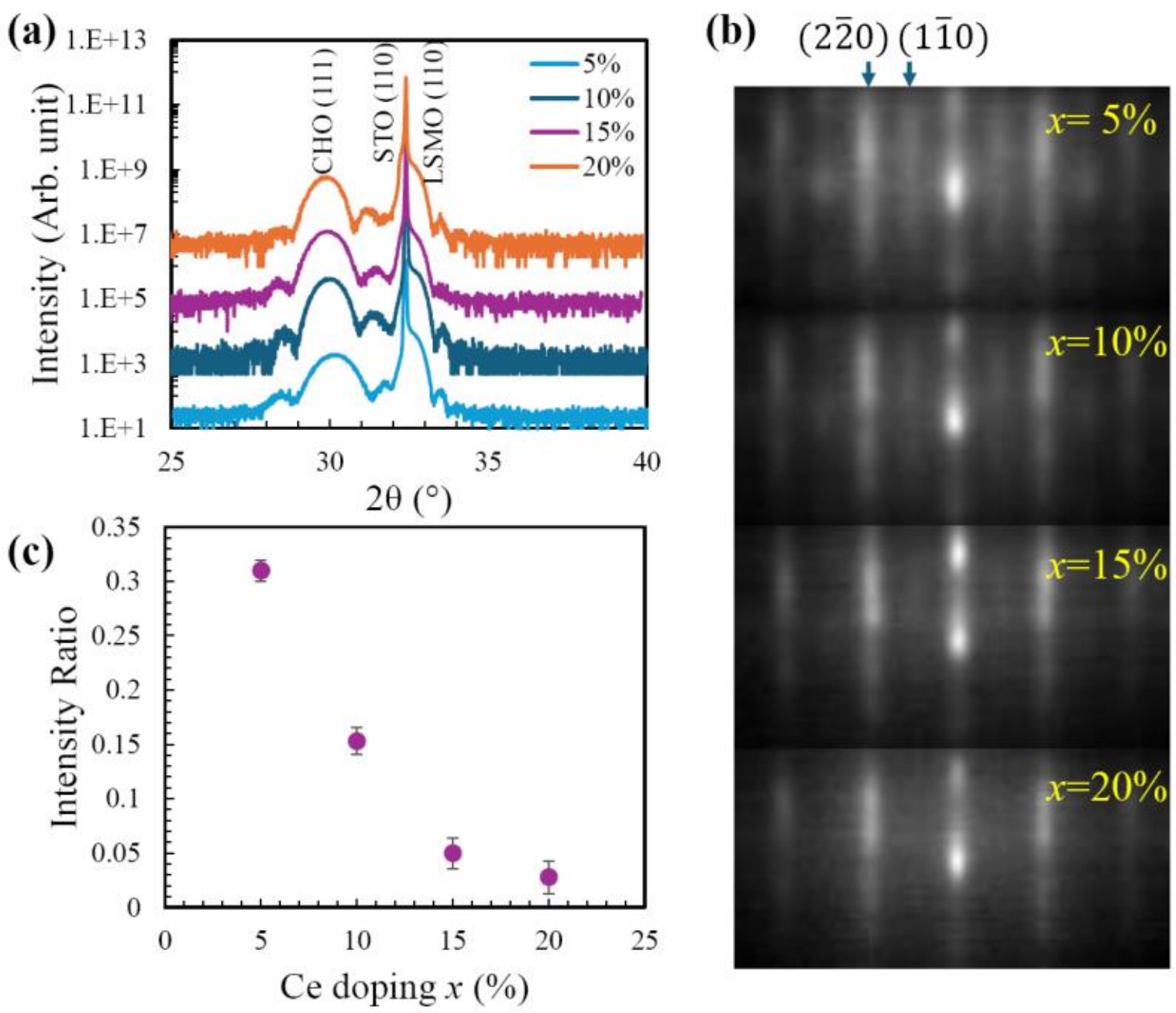

**Figure 1.** (a) θ-2θ x-ray diffraction of CHO (111)/LSMO(110)/STO(110) films of doping level x=5%, 10%, 15%, and 20%. (b) RHEED images of CHO films of different doping level. (c) Intensity ratio between the pseudo cubic $(1\bar{1}0)$ and $(2\bar{2}0)$ reflections as a function of doping level, calculated from the images in (b).

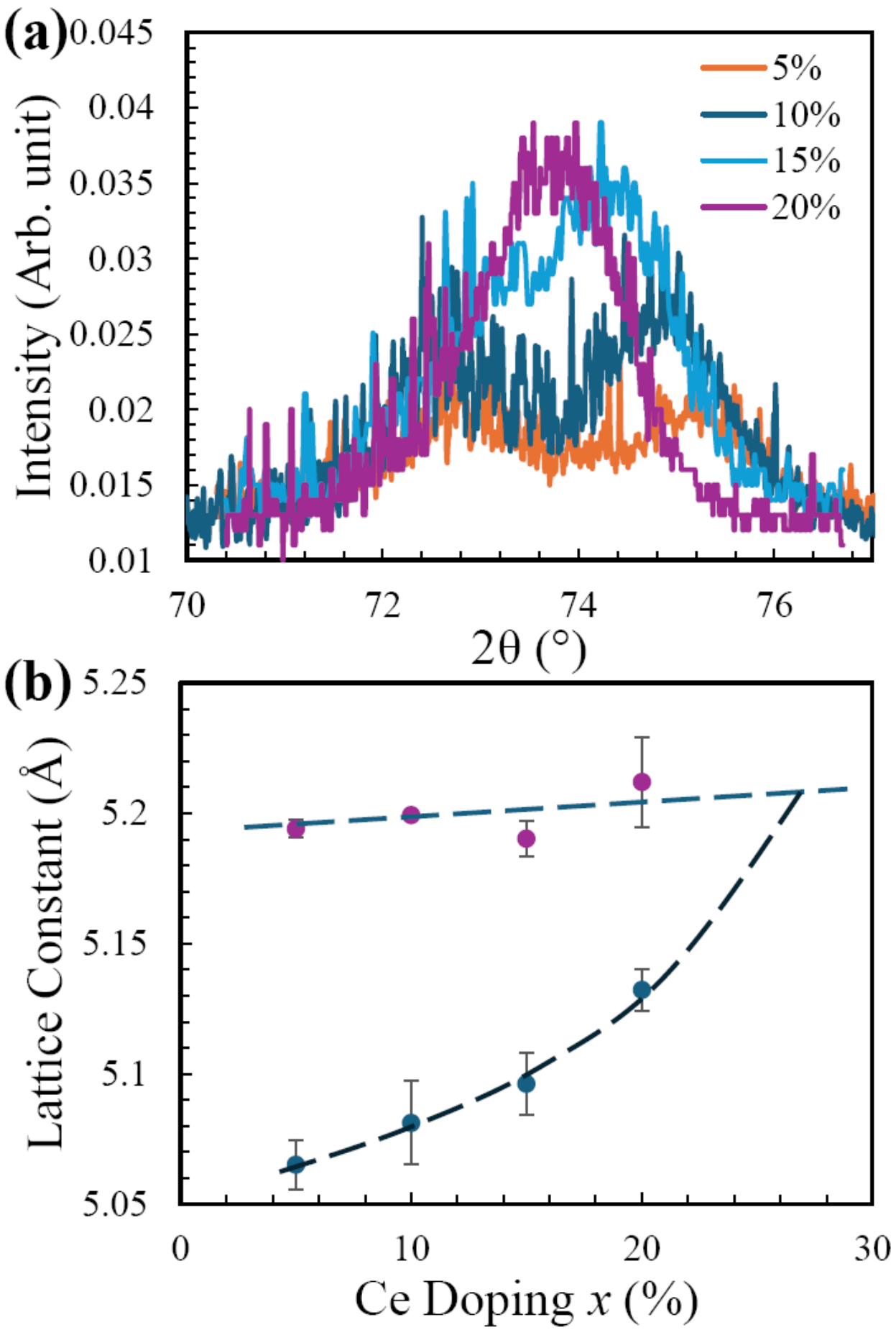

**Figure 2**. (a) X-ray diffraction near the pseudo cubic (400) peaks for different doping levels. (b) Lattice constants extracted from the peak positions in (a); dashed lines are a guide to the eye.

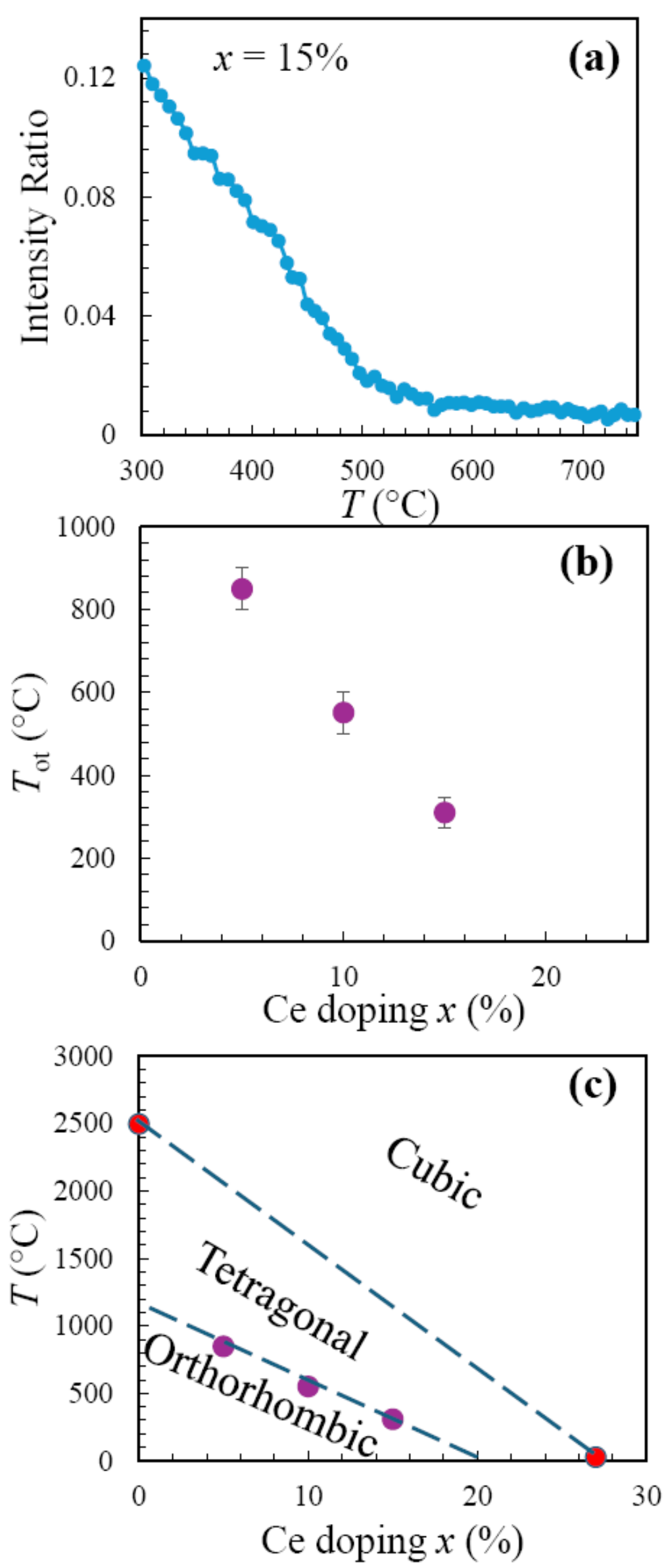

**Figure 3**. (a) Intensity ratio between the pseudo cubic $(1\bar{1}0)$ and $(2\bar{2}0)$ RHEED peaks as a function of temperature for $x$=15%, which indicates a $T_{ot}$ about 550 °C. (b) $T_{ot}$ as a function of doping level. (c) Temperature-doping phase diagram. The dashed lines are a guide to the eye.

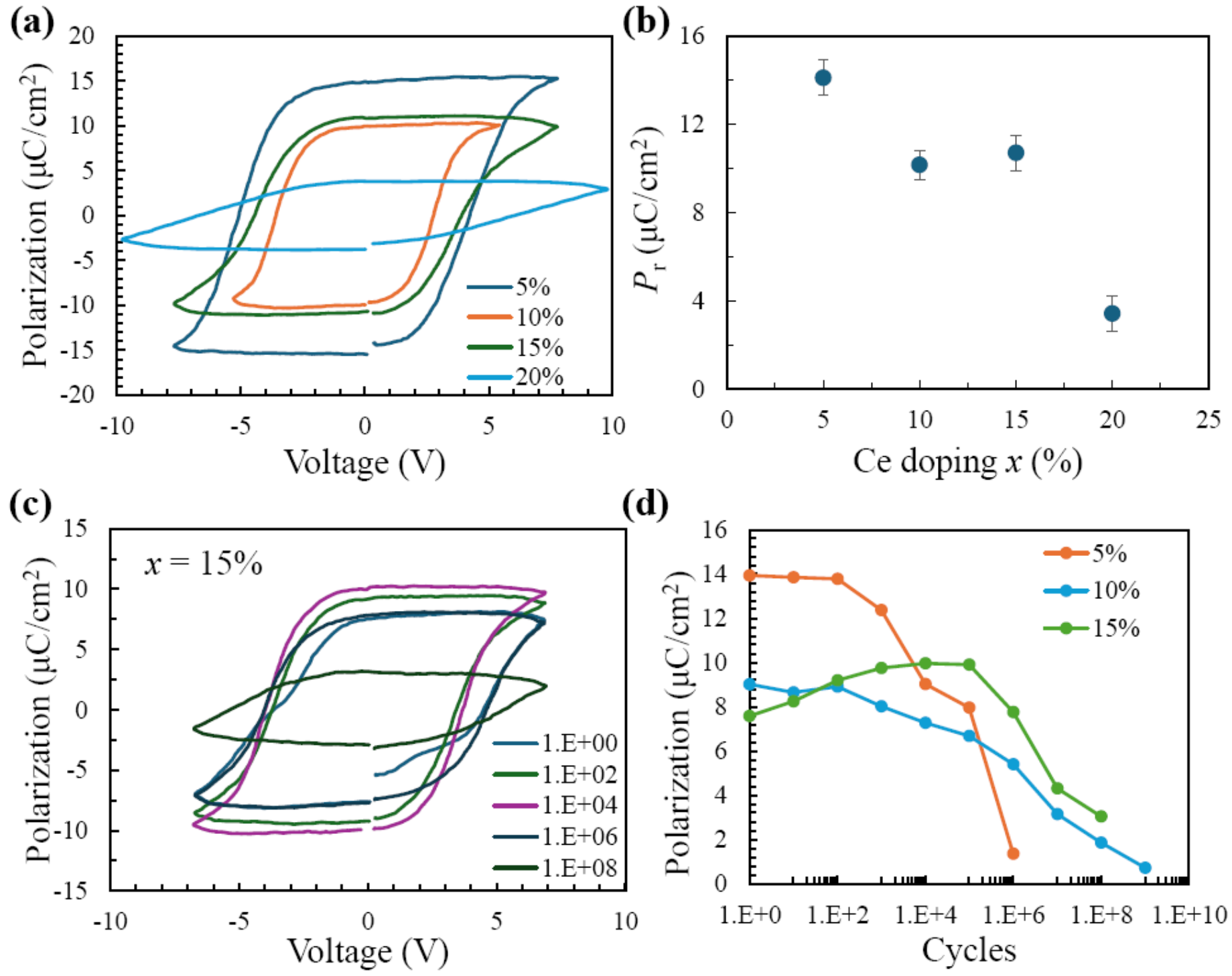

**Figure 4**. (a) Polarization switching loops of CHO (111)/LSMO(110)/STO(110) films of doping level $x$=5%, 10%, 15%, and 20%. (b) Remanent polarization as a function of doping level extracted from (a). (c) Polarization switching loop of a $x$=15% CHO film after up to $10^8$ cycles. (d) Polarization as function of number of switching cycles.